\documentclass[amsmath,amssymb,aps,floats,prx,twocolumn,showpacs,superscriptaddress]{revtex4-1}

\usepackage{graphics}
\usepackage{graphicx}
\usepackage{amssymb}
\usepackage{epstopdf}
\usepackage{color}

\DeclareGraphicsExtensions{.pdf, .png, .jpg, .eps}
\usepackage{floatrow}

\newcommand{\bfig}{\begin{figure}}
\newcommand{\efig}{\end{figure}}

\newcommand{\Tr}{{\text{Tr}}}

\newcommand{\Hcef}{\mathcal H_{\rm CEF}}
\allowdisplaybreaks

\usepackage{float}

\newcommand{\bea}{\begin{eqnarray}}
\newcommand{\ena}{\end{eqnarray}}
\newcommand{\bee}{\begin{equation}}
\newcommand{\ene}{\end{equation}}

\newcommand{\cybs} {CsYbSe$_{2}$}

\newcommand{\ymgo} {YbMgGaO$_4$}
\newcommand{\Jxx} {\mathcal J_{\perp}}
\newcommand{\Jzz} {\mathcal J_z}
\newcommand{\muh} {\mu_0H}
\newcommand{\aperp} {\alpha_{\perp}}
\newcommand{\alz} {\alpha_z}

\newcommand{\inversechi} {\chi^{-1}}
\newcommand{\done} {\Delta_{10}}
\newcommand{\dtwo} {\Delta_{20}}
\newcommand{\dthree} {\Delta_{30}}

\newcommand{\chiab} { \chi_{ab}}
\newcommand{\chic} { \chi_{c}}

\newcommand{\invchiab} { \chi_{ab}^{-1}}
\newcommand{\invchic} { \chi_{c}^{-1}}
\newcommand{\xxz} { \rm{XXZ}}

\newcommand{\kab} { k_{ab}}
\newcommand{\Hpab} { H\parallel ab}
\newcommand{\Hpc} { H\parallel c}

\newcommand\numberthis{\addtocounter{equation}{1}\tag{\theequation}}

\begin{document}



\title {Systematic extraction of crystal electric-field effects and quantum magnetic model parameters in triangular rare-earth magnets}

\author{Christopher A. Pocs}
\affiliation{Department of Physics, University of Colorado, Boulder, Colorado 
80309, USA}

\author{Peter E. Siegfried$^{\dagger}$}
\affiliation{Department of Physics, University of Colorado, Boulder, Colorado 
80309, USA}

\author{Jie Xing}
\affiliation{Materials Science and Technology Division, Oak Ridge National 
Laboratory, Oak Ridge, Tennessee 37831, USA}

\author{Athena S. Sefat}
\affiliation{Materials Science and Technology Division, Oak Ridge National 
Laboratory, Oak Ridge, Tennessee 37831, USA}

\author{Michael Hermele}
\affiliation{Department of Physics, University of Colorado, Boulder, Colorado 
80309, USA}
\affiliation{Center for Theory of Quantum Matter, University of Colorado, 
Boulder, Colorado 80309, USA}

\author{B. Normand}
\affiliation{Paul Scherrer Institute, CH-5232 Villigen-PSI, Switzerland}
\affiliation{Institute of Physics, Ecole Polytechnique F\'ed\'erale de 
Lausanne (EPFL), CH-1015 Lausanne, Switzerland}

\author{Minhyea Lee}
\affiliation{Department of Physics, University of Colorado, Boulder, Colorado 
80309, USA}

\date{\today}

\begin{abstract}
A primary goal at the interface of theoretical and experimental quantum 
magnetism is the investigation of exotic spin states, mostly notably quantum 
spin liquids (QSLs), that realize phenomena including quasiparticle 
fractionalization, long-ranged entanglement, and topological order. Magnetic 
rare-earth ions go beyond the straightforward paradigm of geometrical 
frustration in Heisenberg antiferromagnets by introducing competing energy 
scales, and in particular their strong spin-orbit coupling creates multiple 
split crystal electric-field (CEF) levels, leading to anisotropic effective 
spin models with intrinsic frustration. While rare-earth delafossites have 
a triangular-lattice geometry, and thus have gained recent attention as 
candidates for hosting spin-1/2 QSL physics, the reliable extraction of 
effective spin models from the initial many-parameter CEF spectrum is a 
hard problem. Using the example of \cybs, we demonstrate the unambiguous 
extraction of the Stevens operators dictating the full CEF spectrum of 
Yb$^{3+}$ by translating these into parameters with a direct physical 
interpretation. Specifically, we combine low-field susceptibility measurements 
with resonant torsion magnetometry (RTM) experiments in fields up to 60 T to 
determine a sufficiently large number of physical parameters -- effective 
Zeeman splittings, anisotropic van Vleck coefficients, and magnetotropic 
coefficients -- that the set of Stevens-operator coefficients is unique. Our 
crucial identification of the strong corrections to the Zeeman splitting of 
Kramers doublets as van Vleck coefficients has direct consequences for 
the interpretation of all anisotropic magnetic susceptibility measurements.
Our results allow us to determine the nature and validity of an effective 
spin-1/2 model for \cybs, to provide input for theoretical studies of such 
models on the triangular lattice, and to provide additional materials insight 
into routes for achieving magnetic frustration and candidate QSL systems in 
rare-earth compounds.
\end{abstract}

\maketitle

\section{Introduction}
\label{si}

The quantum spin liquid (QSL) is a nonmagnetic many-body ground state 
in which the spin correlations have long-ranged quantum entanglement
\cite{SavaryReview2016}. Despite many theoretical studies of these 
enigmatic phases, which have served to drive detailed investigations 
of a wide range of magnetic compounds, neither a universally agreed 
QSL phase in a real material nor an unambiguous set of experimental 
QSL criteria has yet emerged. The strong quantum fluctuations responsible 
for producing exotic spin states are a consequence of generalized magnetic 
frustration, which leads to a highly degenerate manifold of competing 
states. Early examples of material realizations of candidate models for 
hosting QSL states were based on geometrical frustration in structures 
with triangular motifs, including kagome \cite{Helton2007}, pyrochlore 
\cite{Gardner2010,JRau2019}, and triangular lattices \cite{Shimizu2003,
Yamashita2011,Itou2008,Itou2010}, mostly of real $S = 1/2$ spins. 

Magnetic insulators with strong spin-orbit coupling are now widely recognized 
as a platform for extending very significantly the nature of frustration and 
the variety of quantum many-body phases (including QSLs) and phenomena that 
can be realized. When compared to spin-1/2 magnetic insulators based on $3d$ 
ions, these systems tend to exhibit complex microscopic physics even at the 
single-ion level. Magnetic materials containing $5d$ and $4d$ transition-metal 
ions possess interactions that are anisotropic in both spin space and real 
space, leading to complex phenomenology in pyrochlore systems 
\cite{Gardner2010,Matsuhira2011,Chen2012} and in proximate 
Kitaev materials \cite{Jackeli2009,Choi2012,Singh2012,Plumb2014,Chun2015}. 
Compounds based on $4f$ rare-earth ions that combine the geometric frustration 
of pyrochlore and triangular lattices with strong spin-orbit coupling have 
also provided fertile ground for quantum magnetism research \cite{Gao2019, 
Mauws2018,Scheie2020,Scheie2021}. However, in these materials a detailed 
understanding of the microscopic physics is a prerequisite for developing 
effective spin models that serve as a basis for theories of many-body 
phenomena.

\begin{figure}[t]
\includegraphics[width=\linewidth]{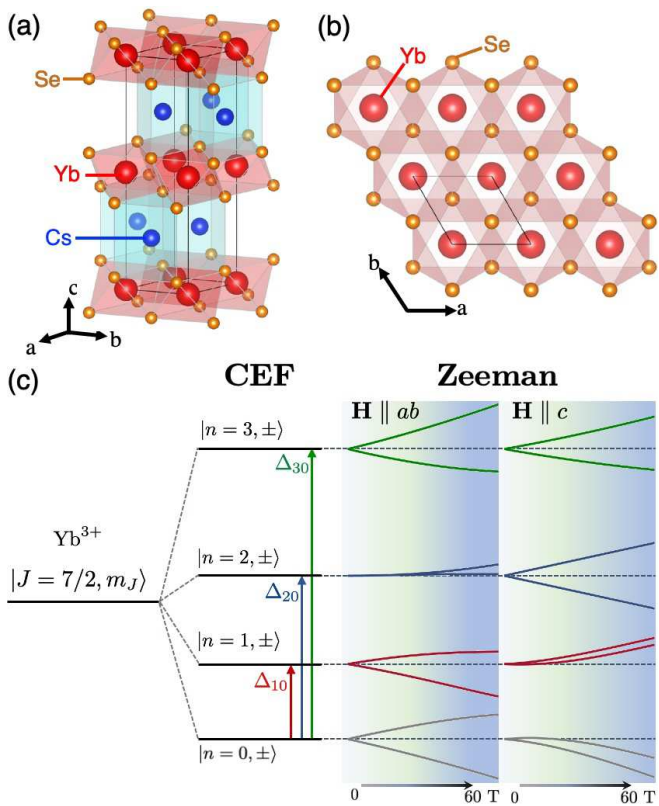}
\caption{Crystal structure and CEF spectrum. (a) \cybs~adopts the 
$P6_3/mmc$ space group. (b) Yb$^{3+}$ (red) triangular layers formed by 
edge-sharing YbSe$_6$ octahedra are separated by layers of Cs$^+$ ions. 
(c) Schematic origin of the CEF energy spectrum of Yb$^{3+}$. The CEF 
interactions allowed by the $D_{3d}$ site symmetry split the Yb$^{3+}$ 
ground-state manifold into four Kramers doublets. The application of a 
magnetic field lifts the doublet degeneracy in a spatially anisotropic 
manner whose leading nonlinear contributions are captured by the van Vleck 
coefficients defined in Eq.~(\ref{eq:vvcoeff}). }
\label{fig1}
\end{figure}

In insulating $4f$ materials, the key to a microscopic description of the 
magnetic interactions is a reliable determination of the single-ion crystal 
electric-field (CEF) Hamiltonian ($\mathcal H_{\rm CEF}$) \cite {Gardner2010,
JRau2019}. In many rare-earth compounds, CEF effects split the degeneracy of 
the ground multiplet of the total angular momentum, $J$. For half-odd-integer 
$J$, the result is a set of Kramers doublets. At sufficiently low temperatures, 
a restriction of the dynamics to the lowest such doublet may be invoked to 
justify using an effective pseudospin-1/2 model \cite{Sibille2015,Gao2019,
Bordelon2019,PDDai2021}. From an experimental standpoint this allows a 
significant diversification of both materials and magnetic phenomena, 
while on the theoretical side a minimal model may be relatively simple, or 
even a realization of one of the paradigm models in frustrated magnetism.  

The leading system-specific parameters in $\Hcef $ are commonly determined 
by inelastic neutron scattering (INS), which provides direct information 
about the spectrum of CEF multiplets. However, there are in general more 
symmetry-allowed parameters in $\Hcef$ than there are gaps between CEF 
multiplets, resulting in an underdetermined fit of the CEF parameters and 
hence to uncertainties in the appropriate pseudospin-1/2 model that can be 
qualitative rather than merely quantitative corrections. Methods allowing 
the unambiguous determination of $\Hcef$ with higher reliability are thus 
very desirable. In this article, we present a high-fidelity determination 
of the CEF parameters of a selected $4f$ system, obtained by using a 
combination of low-field magnetic susceptibility and high-field (up to 
60 T) resonant torsion magnetometry (RTM) measurements.

Yb-based triangular-lattice compounds offer an excellent combination of 
geometric frustration, quasi-two-dimensional nature, half-odd-integer 
$J$, and a strongly split CEF spectrum that suggests the validity of 
pseudospin-1/2 models of quantum magnetism. In particular, the family 
of $A$Yb$X_2$ delafossites (with $A =$ Na, Cs and $X =$ O, S, Se) has 
attracted intensive interest \cite{Baenitz2018,Ranjith2019a,Ranjith2019b,
Sarkar2019,Bordelon2019,Sichelschmidt2020,Bordelon2020,Zhang2021,PDDai2021}, 
and a very recent summary was compiled in Ref.~\cite{Schmidt2021}. Unlike 
the material YbMgGaO$_4$ \cite{Li2015,Shen2016,Li2016,Paddison2017,Li2018,
Shen2018}, they are free from potential site disorder \cite{YLi2017,ZZhu2017} 
and no members of the family have been found to exhibit magnetic ordering at 
temperatures down to tens of mK \cite{Schmidt2021}, while several have been 
reported to exhibit continua of low-energy magnetic excitations 
\cite{Bordelon2019,PDDai2021,Xie2021}. Their magnetic response is highly 
anisotropic between the in- and out-of-plane directions, providing a valuable 
opportunity to use the magnetic field to vary the free-energy landscape. Thus 
the trianguar-lattice delafossites present an ideal test case for unambiguous 
fitting of the field-induced CEF spectra and subsequent establishment of the 
effective spin-1/2 states in rare-earth compounds.

As an example material we focus on \cybs~[Figs.~\ref{fig1}(a-b)]. The complete 
CEF spectrum is specified by the coefficients of six Stevens operators, and 
thus the challenge is to measure enough independent physical quantities beyond 
the low-energy limit. From the susceptibility we extract not only the linear 
Zeeman coefficients for both of the primary, high-symmetry field directions 
but also the van Vleck (VV) coefficients for the ground doublet, which are the 
second-order corrections (i.e.~quadratic in the applied field). While these VV 
coefficients are often used to describe the magnetic response of rare-earth 
compounds in terms of an additive contribution to the susceptibility 
\cite{Banda2018,Ranjith2019b}, in fact they are embedded in a nontrivial way 
in the full expression for this quantity. We will show how the multiple roles 
of the ground-state VV coefficients and their significance as stand-alone 
physical quantities, with direct implications for the high-field magnetic 
anisotropy, have yet to be appreciated in connection with CEF fitting. They 
facilitate the bridge to the field range covered by our RTM measurements, 
from which we extract the full field- and temperature-dependence of the 
magnetotropic coefficients for the same two high-symmetry field directions. 
These round out a complete set of eight observables, allowing us to determine 
a unique set of microscopic CEF parameters and thus the full CEF spectrum. As 
Fig.~\ref{fig1}(c) makes clear, our results reveal an intricate energy 
landscape and level-repulsion behavior in the CEF spectrum of \cybs~up 
to high magnetic field values and for both field directions.  

The physical content of our analysis is to interpret the essential role of 
the VV coefficients in determining magnetic properties, even at low fields 
and temperatures, where they are often neglected in pseudospin-1/2 models. 
Quantifying the VV corrections allows us to demarcate the field-temperature 
range over which a pseudospin-1/2 approximation is appropriate in \cybs, 
and to describe the high-field limit accurately. The resulting full 
characterization of the CEF spectra is essential for investigating 
their experimental consequences, for example when the application of intense 
magnetic fields causes multiple real or avoided level-crossings in narrowly 
spaced CEF spectra. It is also a prerequisite to study further mechanisms 
leading to different forms of magnetic frustration, in which additional 
degrees of freedom, such as phonons, hybridize with the electronic spectrum 
to cause profound effects to appear in low-temperature thermodynamics and 
transport properties \cite{BQLiu2018,Cermak2019}. Finally, while we have 
focused on one example material, our analysis can be applied widely to 
localized $4f$-electron systems.

The structure of this paper is as follows. In Sec.~\ref{smm} we introduce 
\cybs, our experimental methods, the complete Stevens operator formalism for 
CEF levels and some approximate treatments. In Sec.~\ref{sechi} we show all 
the results of our susceptibility measurements and their analysis for the two 
primary field directions. Section \ref{sertm} presents our RTM measurements 
and extraction of the magnetotropic coefficients, allowing us to determine 
the full set of Stevens operators. In Sec.~\ref{sd} we discuss the physical 
interpretation of the results and their consequences for experimental 
analysis, effective pseudospin-1/2 models in frustrated quantum magnetism, 
and materials selection for candidate QSL systems. Section \ref{ss} contains 
a brief summary of our contribution and four Appendices provide additional 
information concerning details of the analysis. 

\section{Materials and Methods}
\label{smm}

While most of the $A$Yb$X_2$ family crystallizes in the $R\bar3bm$ space group 
\cite{Schmidt2021}, the layer stacking sequence of \cybs~gives a $P6_3/mmc$ 
structure \cite{Xing2019,Xing2020acs}, in which the triangular-lattice planes 
are constructed from edge-sharing YbSe$_6$ octahedra, as illustrated in 
Figs.~\ref{fig1}(a-b). The ratio of inter- and intra-layer distances 
separating the Yb$^{3+}$ ions results in a quasi-2D magnetic system 
\cite{Xing2019}. Within each plane, the Se atoms mediate identical 
AFM superexchange interactions between all nearest-neighbor Yb$^{3+}$
ion pairs, resulting in highly frustrated triangular magnetism. In 
fact the system does not order at zero field for temperatures down to 
$0.4$ K, although an applied in-plane field induces an up-up-down 
ordering, as demonstrated by the observation of a 1/3 plateau in the 
magnetization, $M(H)$, at this temperature \cite{Xing2019}.  
 
Magnetization and susceptibility measurements were performed on 
single-crystalline samples using a Quantum Design MPMS. All 
susceptibilities we report are obtained from $\chi_{ab,c} = 
\frac{d M_{ab,c}}{dH}$, where the indices denote measurements 
performed with the field oriented in the triangular-lattice plane 
($H \parallel ab$) or perpendicular to it ($H \parallel c$). 

Beyond the 7 T upper limit of our magnetization measurements, 
we employ resonant torsion magnetometry (RTM) to probe the nature 
of the magnetic anisotropy. The sample is mounted on the tip of 
a vibrating cantilever and the measured shifts in the resonant 
frequency ($f_0 \approx 40$ kHz) reflect changes in the magnetic 
rigidity caused by changes in the direction of the applied magnetic 
field, which are quantified by a tensor of magnetotropic coefficients, 
$k({\vec H})$. We focus on $k(H,T)$ at the high-symmetry angles 
$\theta = \pi/2$ (to measure $\kab$) and 0 ($k_c$), where $\theta$ 
is the polar angle of the applied field measured from the crystalline 
$c$-axis. The measurement configuration is summarized in 
Sec.~\ref{sertm} and described in detail elsewhere \cite{Modic2018,
Modic2020}. All RTM data were taken using the capacitive magnet of 
the NHMFL pulsed-field facility at Los Alamos National Laboratory. 

\subsection{CEF Analysis and Model Hamiltonian}

Yb$^{3+}$ ions subject to the CEF interactions of their surrounding 
anion charge distribution have a total $J = 7/2$ ground-state multiplet 
of allowed electronic states. Unlike the triangular-lattice material 
\ymgo, the Yb$^{3+}$ ions in \cybs~are largely free from any site 
disorder associated with the non-magnetic ions. To describe \cybs, 
we parametrize the single-ion CEF interaction as a linear combination 
of the six symmetry-allowed Stevens operators, $\hat{O}_n^m$, for the 
$D_{3d}$ site symmetry of the YbSe$_6$ octahedral environment
\cite{Baenitz2018,Ranjith2019a,Ranjith2019b,Sarkar2019,Bordelon2019, 
Bordelon2020,Zhang2021,Schmidt2021} to obtain
\begin{equation}
\hat{\mathcal{H}}_{\rm CEF} = B_2^0\hat{O}_2^0 + B_4^0\hat{O}_4^0
 + B_4^3\hat{O}_4^3 + B_6^0\hat{O}_6^0 + B_6^3\hat{O}_6^3
 + B_6^6\hat{O}_6^6  \numberthis
\label{eq:CEF}
\end{equation}
This Hamiltonian splits the $J = 7/2$ multiplet into four Kramers 
doublets, $|n_{\pm} \rangle$ with $n = 0$, 1, 2, 3, whose energies 
we use to define the separations from the ground-state doublet 
($n = 0$) as $\done$, $\dtwo$, and $\dthree$ [Fig.~\ref{fig1}(c)]. 
The corresponding wave functions are obtained by diagonalizing 
Eq.~(\ref{eq:CEF}). A CEF spectrum for the zero-field limit can 
be identified using spectroscopic probes, particularly INS and 
Raman spectroscopy, but to date little information is available 
with which to investigate the high-field reorganization of the 
energy spectrum.  

The symmetries of the triangular lattice of Yb$^{3+}$ ions permit 
a nearest-neighbor superexchange interaction with XXZ spin symmetry 
\cite{Yamamoto2014}. Additional symmetry-allowed and bond-dependent 
anisotropic pseudo-dipolar exchange terms \cite{Maksimov2019} are found 
to give vanishing contributions in a standard Weiss mean-field approximation. 
Thus we restrict our analysis to the minimal XXZ spin model, $\hat{H}_{\rm XXZ}$, 
describing nearest-neighbor interactions between adjacent in-plane $J = 7/2$ 
moments in terms of two interactions, $\Jxx$ and $\Jzz$, which are the 
respective couplings of spin components transverse and parallel to 
$\hat{z}$, i.e. 
\begin{equation}
\hat{\mathcal{H}}_{\rm XXZ} = \sum_{\langle ij \rangle} \! \left[
\Jxx \! \left( \! \hat{J}_{i,x} \hat{J}_{j,x} + \hat{J}_{i,y} \hat{J}
_{j,y} \! \right) \! + \Jzz \hat{J}_{i,z} \hat{J}_{j,z} \right] \! ,
\label{ehxxz}
\end{equation}
in which the indices $\langle i,j \rangle$ refer to nearest-neighbor 
lattice sites and $\hat J_{i,\gamma}$, with $\gamma = x,y,z$, labels 
the components of spin $J = 7/2$ moments on site $i$. 

To account for Zeeman coupling to the external field we add the term
\begin{equation}
\hat{\mathcal{H}}_{\rm Z} = - \mu_0 \mu_B g_J \mathbf{H} \cdot \sum_{i} 
\hat{\mathbf{J}}_i,
\label{eq:zeeman}
\end{equation}
where the Land\'e $g$-factor, $g_J = 8/7$, is used for Yb$^{3+}$ moments. We 
note that the quantization axis defining the $\hat{z}$ direction of the 
chosen basis of spin operators in $\hat{H}_{\rm CEF}$ and $\hat{H}_{\rm XXZ}$ is 
identically the crystallographic $c$-axis, whence the component $H_z$ refers 
to a field ($\mathbf H \parallel c$) applied along the $c$-axis in experiment. 
Similarly, $H_{\perp} = \mathbf{H} \parallel ab$ refers to a field component 
perpendicular to the $c$ axis, which lies precisely in the hexagonally 
symmetric $ab$-plane. In none of our experiments (magnetization, 
susceptibility, RTM) did we find a discernible difference in the response 
for different in-plane field directions, and hence we do not distinguish 
between these. The two separate contributions to the spin response under 
an external magnetic field, arising from the single-ion anisotropy and 
the superexchange anisotropy (both of which have their origin in the CEF 
spectrum), are then captured by the Hamiltonian  
\begin{equation}
\hat{\mathcal{H}}_{\rm tot} = \hat{\mathcal{H}}_{\rm CEF}
 + \hat{\mathcal{H}}_{\xxz} + \hat{\mathcal{H}}_{\rm Z}. 
\label{eq:hamfull}
\end{equation} 

\subsection{Weiss Mean-Field Approximation}

We treat the physics of the system at finite temperature within a 
self-consistent Weiss mean-field approximation for the Yb$^{3+}$ 
spins, which we will find captures the bulk magnetic behavior of 
\cybs~exceptionally well at all measurement temperatures ($T \ge 
2 $ K). In the Weiss mean-field treatment, Eq.~(\ref{eq:hamfull}) 
reduces to a system of $N$ decoupled Yb$^{3+}$ ions each subject 
to the Hamiltonian 
\begin{align*}
\hat{\mathcal{H}}_{\rm MF}^{\rm sg} & = \hat{\mathcal{H}}_{\rm CEF}
 - {\textstyle \frac12} q \left[ \Jxx \langle \hat{J}_x \rangle^2
 + \Jzz \langle \hat{J}_{z} \rangle^2 \right] \numberthis \label{eq:HMF} \\
 & \quad - \left[ \mu_0 \mu_B g_J \mathbf{H} - q \left( \Jxx \langle 
\hat{J}_x \rangle \hat{\mathbf{x}} + \Jzz \langle \hat{J}_{z} \rangle 
\hat{\mathbf{z}} \right) \right] \! \cdot \! \hat{\mathbf{J}},  
\end{align*} 
where $q = 6$ is the nearest-neighbor coordination number on the triangular 
lattice and the mean-field expectation values of the spin operators are 
determined self-consistently as $\langle \hat{J}_\alpha \rangle = \Tr 
[\hat{J_\alpha} e^{-\beta \hat{\mathcal{H}}_{\rm MF}^{\rm sg}}]/Z$, where $Z = \Tr[e^{-\beta 
\hat{\mathcal{H}}_{\rm MF}^{\rm sg}}]$ and $\beta = 1/k_B T$. Because we find 
that the bulk magnetic susceptibility of \cybs~is almost entirely uniaxial, 
we make the additional simplifying assumption that the external magnetic 
field lies in the $xz$-plane, such that $\mathbf{H} = H_\perp \hat{\mathbf{x}}
 + H_z \hat{\mathbf{z}}$.

The precise determination of the mean-field Hamiltonian given in 
Eq.~(\ref{eq:HMF}) poses a major challenge because it contains eight 
unknown parameters, six coefficients $\{B_m^n\}$ of the Stevens operators 
in Eq.~(\ref{eq:CEF}) and two energy scales, $\Jxx$ and $\Jzz$, specifying 
the interactions of the $J = 7/2$ spins. INS spectra are used widely as the 
starting point for extracting the coefficients $\{B_m^n\}$ \cite{Scheie2020,
Scheie2021}, but as noted above measuring the three CEF level-splittings, 
$\done$, $\dtwo$, and $\dthree$ [Fig.~\ref{fig1}(c)], is not sufficient to 
determine six unknowns. Electron spin resonance (ESR) \cite{Sichelschmidt2020} 
probes a much lower frequency range, making it the method of choice for 
determining effective $g$-factor values in the ground doublet, while 
the $T$-dependence of the line width can be used to estimate $\done$
\cite{Baenitz2018,Ranjith2019b,Schmidt2021}; in very well characterized 
systems, the line width may also be used to discuss the effective exchange 
anistropy between $\Jxx$ and $\Jzz$ \cite{Sichelschmidt2020}.

Thus the parameter space for fitting the measured CEF spectra usually remains 
highly degenerate, even with accurate spectral measurements in an applied field 
and well-constrained $g$-tensors, and hence the uniqueness of a fitted set of 
Stevens-operator coefficients cannot be guaranteed \cite{YLi2017,Bordelon2020,
Zhang2021}. The method we apply to solve this problem has two key components. 
First we apply a detailed consideration of the second-order corrections in the 
field-dependence of the CEF spectrum, encoded as the VV coefficients we extract 
from the low-field susceptibility. Then we leverage extensive RTM data 
providing systematic temperature- and field-dependent information about the 
magnetotropic coefficients for the two high-symmetry field directions to fix 
a unique set $\{B_m^n\}$.

\begin{figure}[t]
\includegraphics[width=\linewidth]{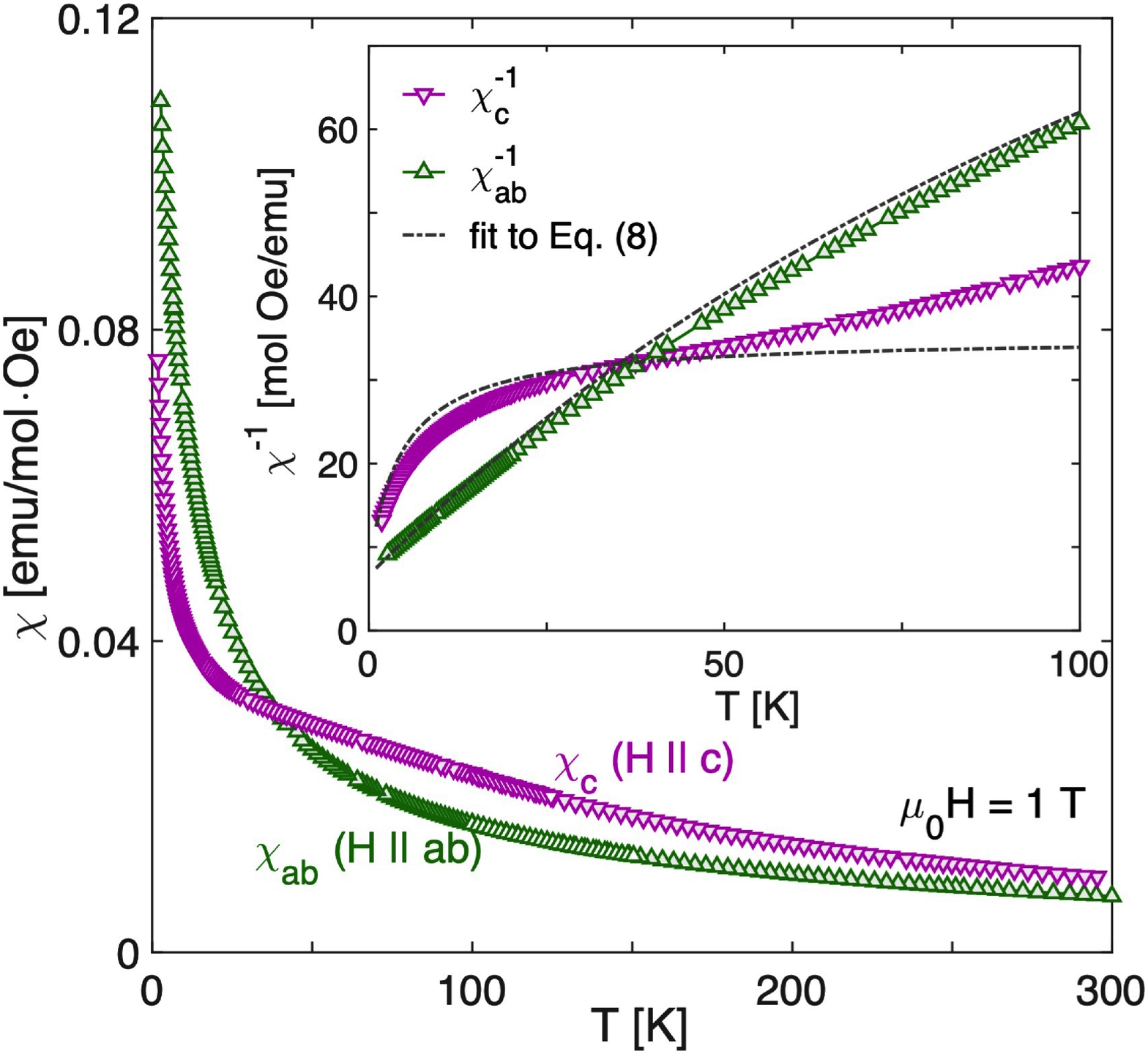}
\caption{Temperature-dependence of the magnetic susceptibilities, $\chi_{ab}$ 
and $\chi_c$, measured for \cybs~in a field of $\muh = 1$ T applied 
respectively in the $ab$ and $c$ directions. The inset shows the inverse 
susceptibilities, $\invchiab$ and $\invchic$, compared with fits obtained 
by applying Eq.~(\ref{eq:vvfit}) in the regime $T \le 45 $ K ($\ll \done/k_B$) 
in order to determine the interaction parameters, $\Jxx$ and $\Jzz$, and the 
VV coefficients, $\aperp^0$ and $\alz^0$, for the ground-state doublet.}
\label{fig:chidata}
\end{figure}

\section{Anisotropic magnetic susceptibilities}
\label{sechi}

\subsection{Experiment: deviations from Curie-Weiss}

In Fig.~\ref{fig:chidata} we show the temperature-dependences of the  
two magnetic susceptibilities, $\chiab (T)$ and $\chic (T)$, measured 
for \cybs~in the presence of an external magnetic field $\muh = 1$ T 
applied respectively within the $ab$ plane and along the $c$ axis. 
There is no indication of long-range ordering down to $T = 2$ K. 
The crossing of the two curves around $T = 35$ K reflects a crossover 
from easy-plane behavior $(\Delta \chi = \chiab - \chic > 0)$ at low 
temperatures to easy-axis anisotropy ($\Delta\chi < 0$) at high $T$. 

Both susceptibilities increase rapidly as the temperature is lowered 
below 50 K, and the corresponding inverse quantities shown in the 
inset of Fig.~\ref{fig:chidata} make clear the susceptibility 
anisotropy, with $\invchic (T)$ exhibiting a significantly sharper 
downward trend as $T$ decreases. While a qualitative inspection 
suggests that a Curie-Weiss form might capture $\invchiab (T)$ for  
$T < 50K$, this is clearly impossible for $\invchic (T)$. This type 
of low-$T$ downturn in $\chi^{-1}(T)$ is observed quite commonly in 
similar rare-earth magnetic materials \cite{Banda2018,Baenitz2018,
Ranjith2019b,Sibille2015,Gaudet2019,Gao2019}, but to date lacks a 
detailed analysis. Next we show that this behavior can be explained 
entirely by analyzing the field-induced evolution of the CEF energy 
levels, where the leading deviations from a linear (effective Zeeman) 
form are contained in two strongly anisotropic ground-state VV 
coefficients.

\begin{table*}[t]
\begin{center}
\caption{Comparison of the lowest zero-field CEF level-splitting, $\done$,
$g$-tensor components, and the VV coefficients for $n = 0$ in a number of 
Yb-based triangular-lattice compounds. For \cybs~we show two sets of 
$g$-factors and VV coefficients, one obtained by a full calculation 
[Eq.~(\ref{eq:vvcoeff})] using the $B_m^n$ coefficients and wave functions, 
i.e.~assisted by fitting to the RTM data as shown in Sec.~\ref{sertm}, 
and the other from fitting $\chi_{ab,c}(T)$ using Eq.~(\ref{eq:vvfit}).
For the other compounds, $\done$ and the $g$-factors are quoted from the 
respective references and we calculated the VV coefficients using the 
reported $B_m^n$ coefficients. The compound nominally most similar to 
\cybs, NaYbSe$_2$, displays slightly less anisotropy in its VV coefficients 
and this is consistent with the differing forms of $\chiab (T)$ and 
$\chic (T)$ reported in Ref.~\cite{Ranjith2019b}. For NaYbO$_2$ we note 
that the two sets of Stevens coefficients (Fit 1 and Fit 2) deduced in 
Ref.~\cite{Bordelon2020} yield dramatically different values of $\aperp^0$ 
and $\alz^0$, which underlines the crucial role of the VV coefficients in 
a complete and consistent characterization of $\Hcef$. YbMgGaO$_4$ is 
found to be least anisotropic among these materials, and its small VV 
coefficients are consistent with the reported Curie-Weiss form of the 
susceptibility.}
\begin{tabular}{ccccccc}
\hline\hline
 &~~$\done$  [meV]~~&~~$g_{\perp}$~~&~~$g_{z}$~~&~$\alpha_{\perp}^0$ 
$[10^{-4} \frac{\rm meV}{\rm T^2}]$~&~$\alz^0$ $[10^{-4}\frac{\rm 
meV}{\rm T^2}]$~& Reference   \\
\hline
\cybs & 13.6 & 3.52 & 1.33 & $-3.31$ & $-18.8$ & calculated from 
Eq.~(\ref{eq:HMF}) 
\\ 
\cybs & $-$ & 3.77 & 1.76 & $-2.93$ & $-17.9$ & $\chi_{ab,c} (T)$ fit to 
Eq.~(\ref{eq:vvfit}) [Fig.~\ref{fig:chidata}]  \\ 
\hline
NaYbSe$_2$ & 17.5/17.7 & 2.87/2.87 & 1.18/1.33 & -5.51/-5.44 & -12.5/-18.8 & 
Fit 1/Fit 2 in Ref.~\cite{Zhang2021} \\ 
NaYbSe$_2$ & 13.8 & 3.13 & 1.01 & $-$ & $-$ & ESR \cite{Ranjith2019b} \\ 
\hline
NaYbO$_2$ & 34.0/34.7 & 3.39/3.54 & 1.71/1.75 &  $-1.10/-2.85$ & $-7.39/-3.62$
& Fit 1/Fit 2 in Ref.~\cite{Bordelon2020} \\ 
\hline
NaYbS$_2$ & 16.7 & 3.19 & 0.57 & - & - & ESR \cite{Baenitz2018} \\ 
\hline
YbMgGaO$_4$ & 39.3/39.4 & 3.22/3.21 & 3.70/3.73 & $-2.25/-2.29$ & $-2.60/-2.58$ 
& Fit 1/Fit 2 in Ref.~\cite{YLi2017} \\ 
\hline
\hline
\end{tabular}
\label{FitVals}
\end{center}
\end{table*}

\subsection{Van Vleck coefficients and low-field limit}

The splitting of the CEF levels in a finite applied magnetic field can be 
understood systematically by perturbation theory. We express the $H$-linear 
and -quadratic corrections to the energy eigenvalues of $\hat{\mathcal 
H}_{\rm CEF}$ due to $\hat{H}_{\rm Z}$ in the form
\begin{align*}
E_{n,\pm} (H) & = E_{n}^0 \pm {\textstyle \frac12} \mu_0 \mu_B \sqrt{ 
(g_\perp^n H_\perp)^2 + (g_z^n H_z)^2} \\ & \qquad + \alpha_{\perp}^n 
H_\perp^2 + \alpha_{z}^n H_z^2 + {\mathcal O}(H^3), \numberthis
\label{eq:energylevel}
\end{align*}
where $E_{n}^0$ refers to the four CEF energy levels at zero field 
(and setting the ground-state energy, $E_0^0$, to zero gives $E_n^0 = 
\Delta_{n0}$ for $n = 1$, 2, and 3). The second term describes the 
$H$-linear Zeeman splitting with a generalized $g$-tensor for all 
levels ($n = 0$, 1, 2, 3) defined as $g_{\perp(z)}^n = 2 g_J \langle 
n_{\pm}| \hat J_{x(z)}|n_{\mp(\pm)}\rangle$, where the subscripts 
$\perp,z$ denote a field applied respectively within the $ab$-plane 
or along the $c$-axis. The conventional effective spin-1/2 $g$-tensor 
components are $g_\perp^0$ and $g_z^0$, to which we refer henceforth, 
without the superscript 0, as the $g$-factors of the system. 

We define all of the phenomena obtained at second order in the perturbative
effect of $\hat{H}_{\rm Z}$ on the zero-field eigenstates of $\hat{H}_{\rm 
CEF}$ as the VV contribution. To describe the full field- and 
temperature-dependence of the susceptibility we define the VV coefficients, 
$\alpha_{\perp}^n$ and $\alpha_z^n$, for the $n$th CEF level by 
\begin{equation}
\alpha_{\perp(z)}^n = (\mu_0 \mu_B g_J)^2 \sum_{n' \neq n} \frac{|\langle 
n'_\pm | \hat{J}_{x(z)} |n_{+} \rangle|^2}{\Delta_{nn'}}, 
\label{eq:vvcoeff}
\end{equation}
where $\Delta_{nn'} = E^0_n - E^0_{n'}$.
Although it is often stated that one may define a ``VV susceptibility,'' 
$\chi_{\rm VV}$, that is a small, additive, temperature-independent, and 
paramagnetic contribution to the total susceptibility, this is rarely an 
accurate approximation. By inspecting the form of the VV coefficients, one
observes that the $n = 0$ terms should be negative, giving the expected 
type of 2nd-order correction to the ground state. The physics content of 
the anisotropic VV coefficients can be read from Fig.~\ref{fig1}(c), where 
the field-dependence of the CEF levels is quite strongly nonlinear due to 
level-repulsion effects between adjacent doublets. In \cybs~this repulsion, 
which is equivalent to a negative curvature of the lower (and positive of 
the upper) branch in each case, is strongest between the $n = 0$ and 1 
doublets for $H \parallel c$.

We obtain analytical formulas for the low-field magnetic susceptibilities 
of the system by using the $N$-particle partition function calculated with 
Eq.~(\ref{eq:HMF}). The full expression for $\chi_{ab(c)} (T)$, which 
includes the contributions of all four CEF levels, is presented in 
Eq.~(\ref {eq:chifull}) of App.~\ref{app_exact}. Here we focus on 
the regime of temperatures sufficiently small that only the lowest-lying 
Kramers doublet need be considered, i.e.~$k_BT \ll \done$, and write 
the inverse susceptibility as 
\begin{eqnarray}
\chi_{ab(c)}^{-1} & = & \frac{1}{\mathcal C} \!\! \left [ \! \frac{T}{|\langle 
0_+ |\hat J_{x(z)}| 0_{-(+)} \rangle|^2 \! - \! \frac{2 k_B \alpha_{\perp (z)}^0}
{\mu_0^2 \mu_B^2 g_J^2} T} + \Theta^{\rm CW}_{\perp(z)} \! \right]  \nonumber \\ 
& = & \frac{\mu_0 k_B}{N} \!\! \left [ \! \frac{T}{{\textstyle \frac{1}{4}} 
g_{\perp(z)}^2 {\mu_0^2 \mu_B^2} \! - \! 2 \alpha_{\perp (z)}^0 k_B T}
 + \Theta^{\rm CW}_{\perp(z)} \! \right] \!\!, \label{eq:vvfit}
\end{eqnarray}
where $\mathcal C = N \mu_0 \mu_B^2 g_J^2/k_B$ is a constant. We have defined 
the quantities $\Theta^{\rm CW}_{\perp(z)} = q \mathcal{J}_{\perp(z)}/k_B$ in order 
to obtain an adapted Curie-Weiss (CW) form for the two applied-field directions 
and in the second line we have used the definition of the $g$-factors (above) 
to make this form more transparent. We note that the regime of validity of 
Eq.~(\ref{eq:vvfit}) is also that in which the system can be approximated by 
an effective pseudospin-1/2 description.

It is clear from Eq.~(\ref{eq:vvfit}) that the contrast to a conventional 
CW form, $\inversechi \propto T + \widetilde\Theta$, is the additional 
$T$-linear term in the denominator, whose prefactor is the corresponding VV 
coefficient. An explicit comparison is presented in App.~\ref{app_spinhalf}. 
The key advantage of our formulation is to observe that the same (VV) 
coefficients describing the $H^2$ correction that gives the leading 
nonlinear contribution to the CEF levels at low temperatures and finite 
fields [Eq.~(\ref{eq:energylevel})] are those describing the nonlinear, 
``beyond-CW'' form of the susceptibility at zero field and finite 
temperatures [Eq.~(\ref{eq:vvfit})]. Thus one may conclude that the latter 
effect is also related to level-repulsion between doublets, the fact that 
the susceptibility is a second field derivative of the free energy meaning 
that second-order perturbative effects do not vanish in the limit $H \to 0$. 

Returning to Fig.~\ref{fig:chidata}, the gray dashed lines in the inset 
show fits to Eq.~(\ref{eq:vvfit}), made in the regime $T < 45$ K, for 
each field direction. From these fits we obtain two exchange energies, 
$\Jxx$ and $\Jzz$ in Eqs.~(\ref{ehxxz}) and (\ref{eq:HMF}), and two VV 
coefficients for the ground-state doublet. A complete fit is deferred 
to Sec.~\ref{sd}. Starting with the VV coefficients, we find the values 
$\aperp^0 = - (2.9 \pm 0.1) \times 10^{-4}$ meV/T$^2$ from $\invchiab$ 
and $\alz^0 = - (17.9 \pm 0.1) \times 10^{-4}$ meV/T$^2$ from 
$\invchic$. We stress that $\alz^0$ is six times larger than 
$\aperp^0$, which for all fields beyond 5 T becomes clearly manifest 
as a much larger downward level-repulsion of the lowest Kramers doublet 
[Fig.~\ref{fig1}(c) and the quantitative analysis of Sec.~\ref{sd}]. 
Turning to the magnetic interactions, we find $\Jxx = 0.54 \pm 0.01$ K
from $\chiab$ and $\Jzz = 0.61 \pm 0.01$ K from $\chic$, both 
encapsulated in $\Theta^{\rm CW}_{\perp(z)}$. If one reduces the system to 
an effective pseudospin-1/2 model, the corresponding interaction terms 
are $\Jxx' = 5.12$ K ($\simeq 0.44$ meV) and $\Jzz' = 0.84$ K ($\simeq 
0.07$ meV), as detailed in App.~\ref{app_proj}. We discuss the physical 
implications of these interaction parameters in Sec.~\ref{sd}B.

As noted above, downward curvature of the low-temperature $\chi^{-1}$ 
has been observed in other rare-earth compounds and our fitting results 
demonstrate that this feature should be characterized by using the VV 
coefficients as a part of a full description of the anisotropic 
magnetism. The validity of Eq.~(\ref{eq:vvfit}) as a replacement 
for the CW form of the susceptibility is confirmed by capturing the 
different $T$-dependences correctly for the two field directions with 
two VV coefficients that are consistent with Eq.~(\ref{eq:energylevel}). 
Although the VV coefficients are not parameters appearing directly in 
the system Hamiltonian, they impose constraints that are essential for 
a unique determination of the full parameter set, a topic we discuss 
further in the next section.

In Table 1 we compare the VV coefficients and other physical characteristics 
for a variety of Yb delafossites. For all compounds listed other than \cybs, 
the $\done$ and $g$-tensor parameters are taken from experimental data. All 
$\aperp^0$ and $\alz^0$ values were calculated from Eq.~(\ref{eq:vvcoeff}) 
using the Stevens coefficients ($B_m^n$) provided by each reference, where 
they were obtained from fits to the CEF spectrum obtained by INS. We stress 
again the fact that, in several of the studies cited, different sets of 
Stevens coefficients can provide equally good descriptions (``Fit 1'' and 
``Fit 2'') of the same INS data due to the underconstrained nature of the 
problem. The anisotropy of \cybs, $\alz^0/\aperp^0 \approx 6$, is strikingly 
higher than the values reported for other compounds in the same family. 
Although all of the NaYbX$_2$ materials seem to show considerable directional 
anisotropy \cite{Schmidt2021}, this may be less severe for NaYbO$_2$, except 
that the two sets of proposed $B_m^n$ parameters yield wildly different VV 
coefficients, indicative of an underlying ambiguity of the type we demonstrate 
how to resolve. In this regard the sole non-delafossite in Table 1, 
YbMgGaO$_4$, is a nearly isotropic outlier. 
 
Before proceeding, we reiterate two important attributes of the VV 
coefficients as a mean of characterizing the magnetic anisotropy of a 
CEF system. First, nonzero VV coefficients are immediately evident in 
$\chi(T)$, as strong deviations from a CW form, and thus failure to 
account for them means that the interaction parameters, $\mathcal 
J_{\perp,z}$, cannot be determined correctly. Second, it is evident 
from Eq.~(\ref{eq:vvfit}) that caution is required in applying an 
effective pseudospin-1/2 description, because even at low $T$, where 
only the lowest Kramers doublet is thermally populated, the repulsion 
from the higher CEF levels cannot be ignored. Hence the twin roles of 
the anisotropic VV coefficients in dictating $H$- and $T$-dependent 
physical properties generic to many $4f$ electronic systems (Table 
1) must be taken into account to obtain a meaningful description of 
the spin physics. 
 
\section{Resonant Torsion Magnetometry}
\label{sertm}

\subsection{Dependence on Field and Temperature}

Having characterized the magnetic response at low fields using $\chi(T)$,
and thereby obtained four independent physical quantities to include in 
the fitting procedure, we turn for more information to the magnetropic 
coefficients. The RTM method allows these to be measured over wide ranges 
of both field and temperature, which we will show provides enough input 
for an unambiguous determination of all the remaining free parameters in 
Eq.~(\ref{eq:HMF}). The magnetotropic coefficient is defined as $k (H,
\vartheta) = \partial^2 F (H,\vartheta)/\partial \vartheta^2$, where 
$F(H,\vartheta)$ is the portion of the Helmholtz free energy depending on 
the magnitude and orientation of the magnetic field ($\vartheta$ is the 
angular direction of ${\hat H}$ measured in the plane of vibration of the 
sample \cite{Modic2018,Modic2020}). $k$ quantifies the magnetic rigidity 
of a material, whose origin lies in the energy cost of rotating a sample 
with an anisotropic magnetic free energy in a finite field, and RTM 
measurements constitute a highly sensitive probe of this magnetoanisotropy. 

The variation of the magnetotropic coefficients at finite field produces 
a shift in the resonant frequency of a system composed of a cantilever 
and an attached sample given by  
\begin{equation}
\dfrac{\Delta f}{f_0} = \dfrac{k}{2K},
\label{eq:fs}
\end{equation} 
with $K$ the intrinsic bending stiffness of the cantilever at zero field 
\cite{Modic2018, Modic2020}. In the low-field limit, where $\chi$ is 
constant at constant temperature, both the torque and the magnetotropic 
coefficient are not only straightforward functions of the polar angle, 
$\theta$, but are also quadratic in $H$. Thus the magnetotropic 
coefficient can be expressed in terms of the difference, $\Delta 
\chi = \chi_{ab} - \chi_c$, between the principal components of the 
susceptibility tensor in the plane of vibration, as $k = \Delta \chi 
H^2 \cos 2 \theta$. More detailed derivations of field-dependent 
expressions for the high-symmetry angles ($\kab$ for $\theta = 0$ 
and $k_c$ for $\theta = \pi/2$) can be deduced by introducing the 
transverse susceptibility, as shown in App.~\ref{app_rtm}.

\begin{figure}[t]
\includegraphics[width=\linewidth]{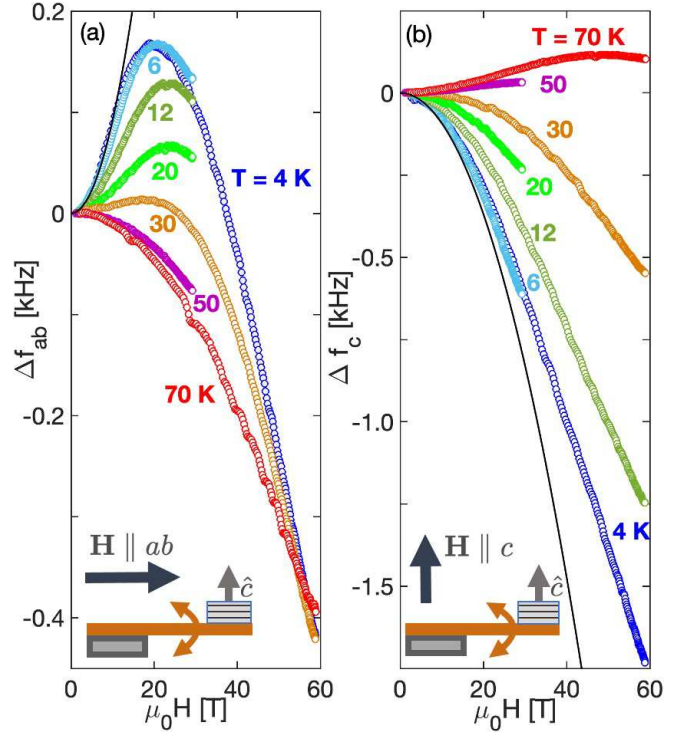}
\caption{Dependence on the applied field magnitude of the resonant 
frequency shifts (a) $\Delta f_{ab}$ for $H \parallel ab$ ($\theta = \pi/2$) 
and (b) $\Delta f_c$ for $H \parallel c$ ($\theta = 0$). Thin solid lines 
show the low-field $H^2$ dependences, with coefficients of $\Delta \chi$ 
and $-\Delta \chi$, taken at $T = 4$ K from Fig.~\ref{fig:chidata}. The 
nonmonotonic $H$-dependence leading to a local maximum in the $H \parallel 
ab$ configuration is attributed to the saturation of the magnetization 
and is explained in Sec.~\ref{sd} by considering the effective 
susceptibility in the transverse direction. The insets illustrate 
the applied-field configurations of the RTM measurements on the 
layered triangular lattice of \cybs~in each case.}
\label{fig:rtmdata}
\end{figure}

Figures \ref{fig:rtmdata}(a) and \ref{fig:rtmdata}(b) show the 
frequency shifts, $\Delta f_{ab}$ and $\Delta f_{c}$, which are 
directly proportional to the magnetotropic coefficients, $\kab$ and 
$k_c$ [Eq.~(\ref{eq:fs})], as functions of field in the $\Hpab$ and 
$\Hpc$ geometries. We note that the vibration plane of the cantilever 
and the rotation plane of $H$ are the same, meaning that in this 
experimental configuration the angle $\vartheta$ of 
Eq.~(\ref{eq:magneotropic}) is identically the polar angle, 
$\theta$, relative to the crystallographic $c$-axis.  
 
At low fields, where $\chi_{ab}$ and $\chi_c$ are $H$-independent constants 
(i.e.~the magnetizations $m_{ab}$ and $m_c$ are linear in $H$), both $k_{ab}$ 
and $k_c$ are indeed proportional both to $H^2$ and to $\Delta \chi$ with 
opposing signs. As the temperature is increased, $\Delta \chi(T)$ in 
Fig.~\ref{fig:chidata} changes its sign above 50 K, and this is reflected in 
the sign-changes of $\Delta f_{ab}$ and $\Delta f_{c}$ between $T = 30$ and 50 K.
 
As the field is increased, $\Delta f_{ab}(H)$ deviates from an $H^2$ 
dependence and becomes nonmonotonic, with a local maximum at low 
temperatures [$T \le 30$ K in Fig.~\ref{fig:rtmdata}(a)]. The origin 
of this behavior lies primarily in $m_{ab}(H)$ saturating in this 
field range, which we confirm from our calculations in Sec.~\ref{sd}. 
By contrast, the behavior of $\Delta f_{c}(H)$ remains monotonic in 
the same $T$ range, where $m_c(H)$ continues to increase with $H$. We 
comment that $\Delta f_c (H)$ does exhibit a weak maximum in $H$ at $T = 70$ 
K, whereas no such behavior is found in $\Delta f_{ab}$ at this temperature. 
This feature is also captured qualitatively by calculating the eigenstates 
of the full mean-field Hamiltonian [Eq.~(\ref{eq:HMF})], as we show next.

\begin{figure}[t]
\includegraphics[width=\linewidth]{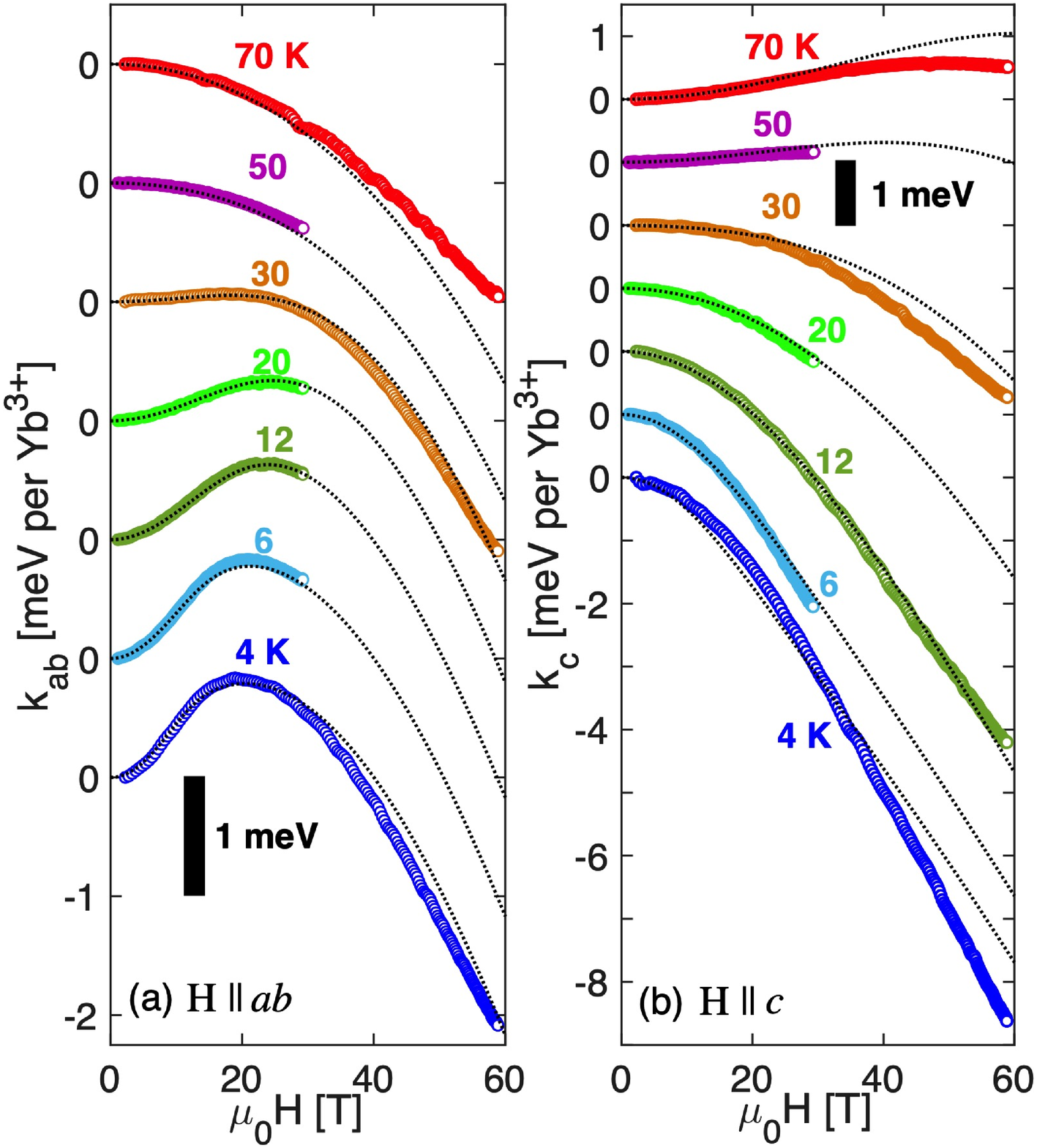}
\caption{Magnetotropic coefficients, $\kab$ and $k_c$, shown per Yb$^{3+}$ 
ion as a function of field in the respective geometries $H \parallel ab$ (a) 
and $H \parallel c$ (b). Open symbols mark measured data and dotted lines 
show the best fits achieved by self-consistent diagonalization of the full 
mean-field Hamiltonian [Eq.~(\ref{eq:HMF})]. The values of $\Jxx$ and $\Jzz$ 
are taken from the susceptibility fits [Fig.~\ref{fig:chifit}], leaving the 
six coefficients of the Stevens operators as free parameters constrained by 
the extracted VV coefficients, $\aperp^0$, and $\alz^0$. Curves are shown 
for clarity with a constant offset.}
\label{fig:rtmfit}
\end{figure}

\subsection{Fitting $\kab$ and $k_c$}

Taking the magnetic interaction parameters $\Jxx$ and $\Jzz$ determined from 
$\chi_{ab,c}(T)$ [Fig.~\ref{fig:chidata}], we fit the measured data for $\kab 
(H)$ and $k_c (H)$ using the full mean-field Hamiltonian based on 
Eq.~(\ref{eq:HMF}), in which the six coefficients of the Stevens operators 
are free parameters to be determined. However, obtaining the ground-state 
VV coefficients, $\aperp^0$ and $\alz^0$, from the fit to Eq.~(\ref{eq:vvfit}) 
reduces the number of fitting parameters to four. Given the wide ranges of 
$T$ and $H$ covered by the RTM data, the remaining unknowns can be determined 
with an unprecedentedly high level of confidence by using $\kab$ and $k_c$.
  
\begin{table}[b]
\begin{center}
\caption{\small Values of the Stevens-operator coefficients 
[Eq.~(\ref{eq:CEF})] obtained from the fits shown in Fig.~\ref{fig:rtmfit}.}
\begin{tabular}{ccccccc}
\hline\hline
unit  &$B_2^0$ & $B_4^0$ &$B_4^3 $& $B_6^0$ &$B_6^3$ & $B_6^6$   \\
\hline
$10^{-2}$ meV & -42.33 & 1.17 & 54.94 & 0.03 & 0.52 &-0.04\\
\hline\hline
\end{tabular}
\label{bnmval}
\end{center}
\end{table}

Figure \ref{fig:rtmfit} displays the magnetotropic coefficients 
converted from the measured $\Delta f$ data in both geometries. 
It is clear that the fits capture the full field-dependence of $\kab (H)$ 
and $k_c (H)$ in an excellent manner, despite the minimal spin model and 
the mean-field approximation. In particular, the low-field $H^2$ curvature 
is fully consistent with $|\Delta \chi|$ (Sec.~\ref{sertm}A). At $\muh > 
30$ T, beyond the range of some of the data, the agreement is no longer 
quantitative for all temperatures simultaneously, but the model continues
to capture the majority of the field-dependent behavior for both field 
directions. One possible source of these deviations would be high-field 
magnetostrictive effects, which can distort the lattice structure and thus 
modify both the magnetic interactions and possibly even the sizes of the 
CEF coefficients at sufficiently large applied fields \cite{Doerr2005}. 
We use the optimal fits to obtain the six Stevens-operator coefficients 
[Eq.~(\ref{eq:CEF})] shown in Table 2 and we discuss the physical 
implications of having this fully determined spin Hamiltonian 
[Eq.~(\ref{eq:HMF})] in the next section.
 
\begin{figure}[t]
\includegraphics[width=\linewidth]{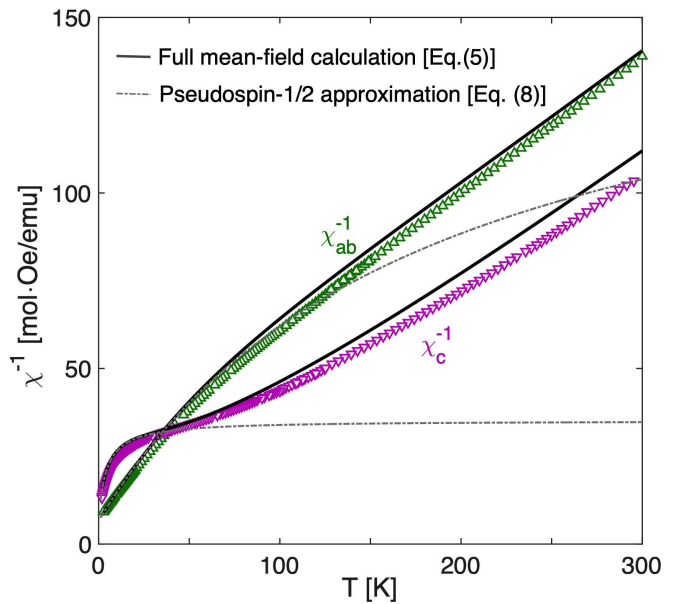}
\caption{Data for the measured inverse susceptibilities (symbols, inset 
Fig.~\ref{fig:chidata}) shown over the full range of temperature together 
with fits (solid lines) computed using the Stevens-operator coefficients 
of Table 2. Thin dot-dashed lines show fits obtained from a pseudospin-1/2 
description [Eq.~(\ref{eq:vvfit})] applied at low temperatures, which align
well with the full matrix calculation for $T < 60$ K.}
\label{fig:chifit}
\end{figure}

\section{Discussion} 
\label{sd}

\subsection{CEF spectrum and VV coefficients}

Having determined a full set of eight fitting parameters from the RTM data, 
we first verify the self-consistency of this fit against the measured 
magnetic susceptibilities, which are shown as $\invchiab(T)$ and $\invchic 
(T)$ over the full temperature range in Fig.~\ref{fig:chifit}. The solid 
lines show the same quantities calculated from the mean-field Hamiltonian 
of Eq.~(\ref{eq:HMF}). The agreement of the fit with the data is 
quantitatively excellent over the entire measured $T$ range, including 
temperatures allowing significant population of the higher CEF levels. 
The dot-dashed lines were calculated using the pseudospin-1/2 approximation, 
i.e.~the contributions from the ground-state doublet [Eq.~(\ref{eq:vvfit})], 
which as expected captures the $T$-dependence below a specific energetic 
cut-off; by inspection we find this cut-off at $T \approx 60$ K in 
\cybs~for both directions of ${\hat H}$.  

As discussed in Sec.~\ref{sechi}, the difference in $T$-dependence for the 
two primary field directions is accounted for in the modelling procedure 
by the large discrepancy in the VV coefficients. While this anisotropy is 
also reflected in the very different prefactors of the $H$-quadratic part 
of the energy spectrum for the two different field geometries, we stress 
again the fact that it affects the susceptibility strongly even at zero 
field. As a consequence, the susceptibility fits in Fig.~\ref{fig:chidata} 
provide an independent determination of $\aperp^0$ and $\alz^0$, as well 
as of the squared matrix elements $|\langle| 0_\pm|\hat J_{x(z)}|0_{\mp 
(\pm)}\rangle |^2$. These constitute additional constraints on the allowed 
Stevens-operator coefficients that are crucial for reducing the enormous 
degeneracy of the six-variable parameter space describing the CEF levels. 
It is well known that suitable sets of coefficients are often highly 
degenerate, in the sense of yielding many indistinguishable fits of 
INS data for the CEF spectrum and $g$-tensor values \cite{YLi2017,
Bordelon2020,Scheie2020}. In Table 1, we compare the ground-state 
$g$-tensor values and VV coefficients obtained by fitting $k_{ab,c} 
(H)$ (top line, Fig.~\ref{fig:rtmfit}) and $\inversechi$ (second line, 
Fig.~\ref{fig:chifit}), where we find agreement at the 10\% level. 

\begin{figure}[t]
\includegraphics[width=\linewidth]{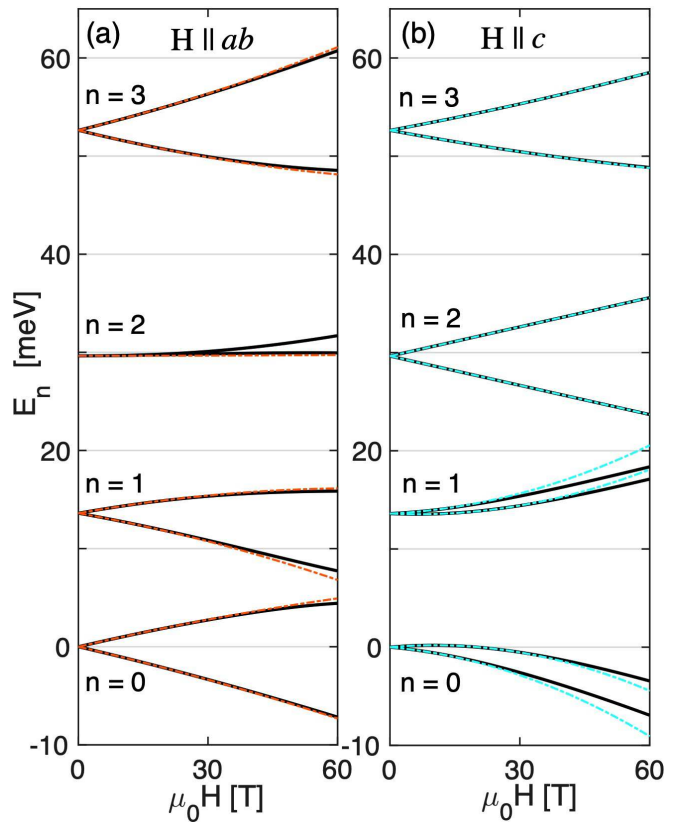}
\caption{CEF spectrum calculated using the fitting result from 
Fig.~\ref{fig:rtmfit}] (solid lines). Quadratic and higher-order 
dependences on the applied field are clearly visible for all four 
Zeeman-split Kramers doublets. We comment that the $n = 2$ level 
remains mostly degenerate in fields $\mathbf H \parallel ab$ 
because of their dipolar-octupolar nature. Dot-dashed lines show  
the perturbative expression of each energy level at second order, 
as in Eq.~(\ref{eq:energylevel}).}
\label{fig:spec}
\end{figure}

Next we use our fit coefficients to calculate the CEF energy levels, 
which are shown in Fig.~\ref{fig:spec} as a function of field at 
zero temperature. The black solid lines show the spectra calculated, 
for ${\hat H}$ oriented in the $ab$-plane [panel (a)] and along the 
$c$-axis [panel (b)], by the direct diagonalization of $\hat{
\mathcal{H}}_{\rm CEF} + \hat{\mathcal{H}}_{\rm Z}$. Focusing first 
on zero field, we obtain the level-splittings $\done = 13.6$ meV, 
$\dtwo = 29.6$ meV, and $\dthree = 52.6$ meV. Thus $\done$ in 
\cybs~is similar to that of NaYbSe$_2$ ($\approx 17.7$ meV 
\cite{Zhang2021}), but is significantly lower than the values 
reported for delafossites in which the Yb$^{3+}$ ion has an oxygen 
environment \cite{Bordelon2020,YLi2017}. The $g$-factor values we 
obtain for the ground state, $g_{ab} = 3.52$ and $g_c = 1.33$, are 
comparable to those measured in all of the Yb-based delafossites 
\cite{Schmidt2021}, as summarized in Table 1. 

Turning to finite fields, a comparison of the ground-state doublet 
($n = 0$) in Figs.~\ref{fig:spec}(a) and \ref{fig:spec}(b) shows 
the expected strong anisotropy, which lies predominantly in the fact 
that the $H$-quadratic component is much larger for $\Hpc$ (the 
aforementioned factor-6 difference between $\alz^0$ and $\aperp^0$).  
Because the quadratic curvatures are a consequence of repulsion 
between adjacent CEF levels, the $n = 1$ doublet exhibits a clear 
opposing curvature for $H \parallel c$ (dominating the behavior of 
both doublet components), whereas in the $\Hpab$ configuration this 
effect is weaker than the higher-order corrections. The dot-dashed 
lines in Fig.~\ref{fig:spec} illustrate the fidelity of a fit made 
only at the level of second-order energy corrections for each $n$ 
[Eq.~(\ref{eq:energylevel})], i.e.~by using all the VV coefficients, 
which agrees well until $\muh > 50$ T in both ${\hat H}$ directions. 

Moving up in the CEF spectrum, our results also capture the unique 
properties of the $n = 2$ state. This dipolar-octupolar doublet, 
$|J, m_J = \pm 3/2 \rangle$ \cite{Huang2014,Sibille2015,Gaudet2019,
YDLi2017}, does not transform as a magnetic dipole because the 
threefold rotational symmetry excludes any mixing between states 
with $m_J = \pm 3/2$ and those with other values of $m_J$ 
\cite{DOnote}. Instead this doublet combines the features of dipolar 
and octupolar moments \cite{Sibille2015,Gaudet2019}. In particular, 
only one component of the dipole moment (in this case oriented along 
the $c$-axis) appears in the vector of pseudospin-1/2 operators, 
which makes the Zeeman splitting of the $n = 2$ level appear very 
different between $\Hpab$ and $H \parallel c$. In the latter case, the 
$m_J = 3/2$ state remains an exact eigenstate up to arbitrarily large 
field, so the Zeeman splitting is exactly $H$-linear with no higher-order 
perturbative corrections [Fig.~\ref{fig:spec}(b)], i.e.~$\alz^{2} = 0$. By 
contrast, for $\Hpab$ the matrix elements $\langle 2_\pm| J_x |2_\mp \rangle$ 
are zero and the $H$-linear Zeeman splitting vanishes $(g_{\perp}^2 = 0)$. Thus 
these doublet components remain nearly degenerate [Fig.~\ref{fig:spec}(a)] 
until the field is sufficiently strong that higher-order perturbative 
corrections become visible. 

\begin{figure}[t]
\includegraphics[width=\linewidth]{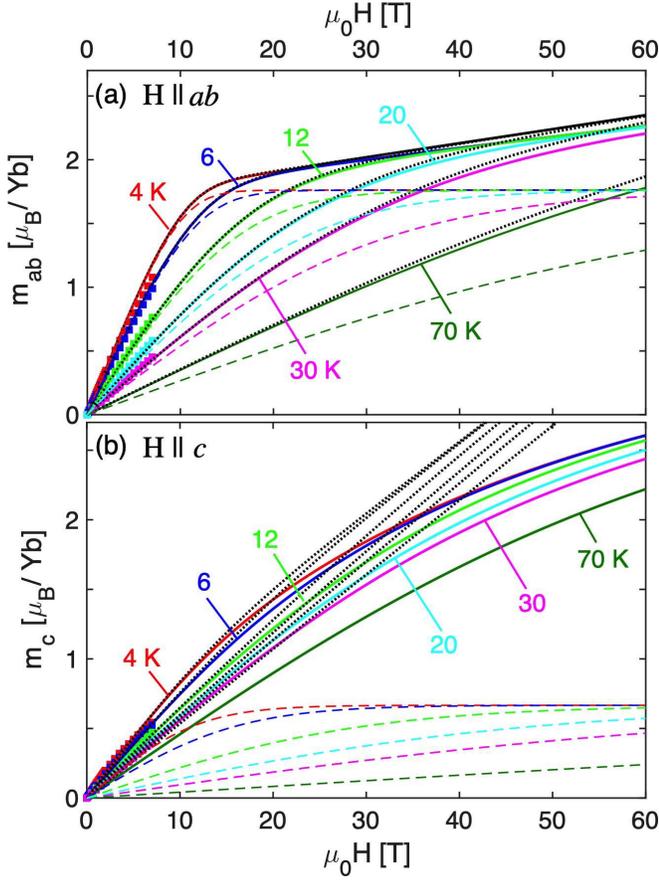}
\caption{Magnetizations, $m_{ab}$ (a) and $m_c$ (b), calculated for selected 
values of the temperature from Eq.~(\ref{eq:HMF}) (solid lines), in a 
pseudospin-1/2 approximation with VV corrections (dotted lines), and in a
pseudospin-1/2 approximation without VV corrections (dashed lines). Solid 
symbols show magnetization data measured up to 7~T.} 
\label{fig:MvH}
\end{figure}

A further key experimental quantity we consider is the magnetization. The 
solid lines in Fig.~\ref{fig:MvH} show the magnetization of the system at 
several different temperatures calculated for fields in the two primary 
directions using the full Hamiltonian matrix [Eq.~(\ref{eq:HMF})]. In the 
low-field range, $m_{ab}$ and $m_c$ reproduce exactly our experimental data 
measured up to $\muh = 7$ T. At higher fields, $m_{ab}$ calculated from the 
full spectrum changes slope at $\muh \approx 12$ T (at $T = 4$ K), beyond 
which it continues to increase more slowly with increasing field. This hint of 
saturation behavior is consistent with the broad maximum in the RTM frequency 
shift, $\Delta f_{ab}(H)$, at low temperatures (Fig.~\ref{fig:rtmdata}). No 
such behavior is found in $m_c$, although the overall slope does decrease as 
$H$ is raised beyond approximately 25 T, and nor is it immediately evident 
in the RTM measurements; however, it can be found within the detailed 
$H$-dependence of the free energy, which we parameterize using the 
transverse susceptibilities, $\chi_{ab}^T (T)$ and $\chi_c^T (T)$, 
in App.~\ref{app_rtm}. 

Finally, the magnetization presents an excellent test case for the validity 
of a pseudospin-1/2 description of \cybs, meaning the extent to which any 
physical property is explained by the contributions from the ground-state 
doublet ($n = 0$) alone. In Figs.~\ref{fig:MvH}(a) and \ref{fig:MvH}(b) 
we show in addition the magnetizations obtained using a pseudospin-1/2 
approximation with (dotted lines) and without (dashed lines) the VV 
correction. For reference, the efficacy of the VV corrections in reproducing 
the full field-induced evolution of the ground doublet can be gauged in 
Fig.~\ref{fig:spec}. For $m_{ab}(H)$, a pseudospin-1/2 model with no VV 
correction captures the field-dependence only up to $\muh \approx 10$ T before 
saturating to the generic tanh function of a Zeeman doublet. By contrast, 
a model with VV correction follows the full solution very closely over the 
entire field range to 60 T. This contrasts with the situation for $m_c (H)$, 
where even the VV-corrected model shows a clear departure from the full 
solution that sets in around 15 T, beyond which the approximate treatment 
separates systematically, and it is clear that higher-order corrections to 
the doublet spectrum are required. For the model with no VV correction, 
$m_c(H)$ does not even reproduce the low-field regime. Recalling the very 
significant magnetic anisotropy ($\alpha^0_z/\alpha_{\perp}^0 \approx 6$), 
the effectiveness of the VV corrections for the two directions is no 
surprise. As a function of temperature, we observe that the VV-corrected 
magnetizations in both geometries show increasing discrepancies at the 
same field as $T$ is increased to 70 K (Fig.~\ref{fig:MvH}), even though 
the nominal thermal cutoff imposed by the next CEF level is $\done/k_B 
\approx 160$ K. 

Thus our experiments and analysis demonstrate two key results. The first 
is that all of the nontrivial magnetic properties of a material are 
captured by a single concept, the VV coefficients. The second is that 
the contrast between conventional and nontrivial magnetic properties can 
be found in a single material due to anisotropic VV coefficients that 
differ strongly between the field geometries $H \parallel ab$ and $H 
\parallel c$. We have shown that the origin of the VV coefficients lies 
in the large matrix elements governing the mixing and consequent repulsion 
of adjacent CEF levels. Although the ground-state VV coefficients, $\aperp^0$ 
and $\alz^0$, are zero-field quantities whose effects can be found in non-CW 
behavior of the susceptibility, their strongest impact emerges as $H$ 
increases (Fig.~\ref{fig:spec}). Our analysis also demonstrates that the 
success or failure of the popular pseudospin-1/2 approximation to the 
magnetic properties of a Kramers-doublet system depends directly on the 
magnitude of the VV coefficients. 

Our conclusions rely completely on determining all of the CEF parameters 
with high accuracy, which allows a detailed characterization of all the 
magnetic properties throughout the $(H,\theta,T)$ parameter space within 
a single and self-consistent model. The simplest indication for a 
significant contribution from the VV coefficients is a deviation of 
the susceptibility from a CW form, which is visible most clearly as a
downward curvature in $\chi^{-1} (T)$ as $T \rightarrow 0$; this type 
of behavior is obvious in $\invchic (T)$ in Fig.~\ref{fig:chifit}, but 
not in $\invchiab(T)$. As noted above, a more detailed characterization 
of the VV coefficients requires experiments under significant applied 
fields. Quantitatively, the origin of the VV coefficients in second-order 
perturbation theory [Eq.~(\ref{eq:vvcoeff})] gives them a systematic 
dependence on $\Delta_{n0}^{-1}$, as well as on the expectation values 
of $\hat J_{x(z)}$. Although the latter terms are primarily responsible 
for the strong directional anisotropy in the Yb-based delafossites whose 
reported $\done$ values and extracted VV coefficients are collected in 
Table 1, the related material YbMgGaO$_4$ presents an example where the 
large value of $\done$ suppresses the VV coefficients, leading to a 
predominantly CW-type behavior of $\chi^{-1}(T)$ \cite{Paddison2017}.

We stress again that our full microscopic model is directly applicable 
to the computation of all aspects of the magnetic response of a material. 
Here we have illustrated the situation for our own magnetic susceptibility, 
magnetization, and RTM data, and we await its extension to describe 
magnetic specific-heat, torque, ESR, and magnetocalorimetric measurements. 
We have also distilled the properties of the full model to the physically 
relevant VV coefficients, which can be understood as governing the  
nonlinear field-dependence of the CEF energy levels (Fig.~\ref{fig:spec}) 
and thus affect all of the magnetic properties. This linking function 
allows the use of our analysis to resolve a number of contradictions that 
have emerged in recent studies of delafossites by different techniques. 

One example is the report from single-crystal INS measurements on \cybs~of 
no CEF levels below 20 meV \cite{Xing2019}, in direct contradiction to the 
present conclusions (Fig.~\ref{fig:spec}). We have calculated the scattering 
cross-section for the $0 \to 1$ transition by using the diagonalized CEF 
matrix in order to compare the intensity obtained in the crystalline geometry 
of the measurement with that expected for a polycrystalline sample. Indeed 
we find the former to be smaller than the latter by four orders of magnitude,  
indicating that the initial conclusion was the consequence of an unfortunate 
choice of geometry, and subsequent reports suggest that a CEF level has been 
found around 15 meV \cite{pcsn}. We note also that the two sets of 
Stevens-operator coefficients proposed \cite{Bordelon2019} for NaYbO$_2$ 
based on INS measurements of the CEF spectrum have widely divergent VV 
coefficients, whereas similar fits to the INS spectra of NaYbSe$_2$ show 
much more agreement when refined with complementary Raman scattering data 
\cite{Ranjith2019b}.

\subsection{Magnetic interaction parameters and triangular-lattice spin models}

We turn our discussion from the Stevens operators determining the CEF levels 
to the magnetic interaction parameters that govern the low-energy physics, 
and hence the extent to which the Kramers-doublet system can be used as an 
effective realization of any of the paradigm $S = 1/2$ models in quantum 
magnetism. As Sec.~\ref{sd}A made clear, an adequate understanding of the 
CEF energies and their evolution in an applied field is a prerequisite in 
the search for phenomena including field-induced phase transitions and 
candidate QSL phases \cite{SavaryReview2016}. In particular, narrowly 
spaced CEF levels that undergo significant mutual repulsion at a specific 
field scale offer a complex and correlated energy landscape that could 
accommodate unconventional spin states. 

Focusing on the triangular geometry, the nearest-neighbor triangular-lattice 
Heisenberg model is the original \cite{Anderson1973} and still one of the 
deepest problems in frustrated quantum magnetism \cite{Li2020,Starykh2015}.
Although the ground state of this model has a modest amount of magnetic order, 
triangular lattices have attracted extensive interest from a number of angles 
over the decades. Not only does this geometry have multiple materials 
realizations, but each generation of materials has opened a new dimension in 
research \cite{Li2020}. Triangular organic compounds drove a discussion of 
proximity to the Mott transition \cite{Shimizu2003,Itou2008,Itou2010,
Yamashita2011}, Cs$_2$CuCl$_4$ and Cs$_2$CuBr$_4$ \cite{Coldea2001,Coldea2002,
Coldea2003,Foyevtsova2011,Zvyagin2014} drove studies of the spatially 
anisotropic ($J$-$J^\prime$) triangular-lattice models \cite{Chen2013,
Starykh2015}, and the cobaltates Ba$_3$CoNb$_2$O$_9$ \cite{Lee2014,
Yokota2014}, Ba$_3$CoSb$_2$O$_9$ \cite{Shirata2012,Ma2016,Ito2017}, and 
Ba$_8$CoNb$_6$O$_{24}$ \cite{Cui2018} spurred the consideration of 
spin-anisotropic triangular lattices with XXZ symmetry. In recent years, 
Yb-based triangular-lattice materials have sparked very strong interest 
in further spin anisotropies, in the form of $J^{++}$ and $J^{+z}$ terms 
\cite{ZZhu2017,Zhu2018,Maksimov2019}, all of which widen considerably the 
scope for finding QSL states. On the theoretical side, it has been shown 
that second-neighbor Heisenberg interactions also drive a QSL state, whose 
gapped or gapless nature remains undetermined at present \cite{Kaneko2014,
Zhu2015,Iqbal2016,Hu2015,Hu2019}. In the nearest-neighbor model, the 
systematic treatment of parton-based formulations has led to qualitative 
advances in calculating the dynamical excitation spectrum by both 
Schwinger-boson \cite{Ghioldi2018} and pseudofermion methods \cite{Iqbal2016,
Ferrari2019}. Long-standing questions about the thermodynamic properties may 
soon be answered by DMRG methods \cite{Chen2019}, despite the constraints of 
working on a rather narrow cylinder, and by tensor-network methods 
\cite{Czarnik2019} despite the challenge posed by the high connectivity 
of the triangular lattice. 

In Sec.~\ref{smm} we restricted our considerations to a triangular-lattice 
model of XXZ form [Eq.~(\ref{ehxxz})], meaning that we allowed only a minimal 
spin anisotropy of Ising or XY form. Working at the mean-field level
[Eq.~(\ref{eq:HMF})], in Sec.~\ref{sechi}B we obtained the results $\Jxx = 
0.54 \pm 0.01$ K and $\Jzz = 0.61 \pm 0.01$ K for the magnetic interactions 
in the system of $J = 7/2$ Yb$^{3+}$ ions. As shown in App.~\ref{app_proj}, 
the interaction terms of the corresponding effective pseudospin-1/2 model are 
$\Jxx' = 5.12$ K ($\simeq 0.44$ meV) and $\Jzz' = 0.84$ K ($\simeq 0.07$ meV), 
meaning that \cybs~is very strongly in the XY limit ($\Jzz'/\Jxx' \simeq 
0.16$). Although this contrasts with most assumptions made at present, we note 
that XY character has also been deduced for each of the NaYb$X_2$ materials by 
using a pseudospin-1/2 treatment \cite{Schmidt2021}. We expect further 
investigations of \cybs~to confirm this result. As the most straightforward 
indicator of XY or Ising physics, we suggest that the width in field of the 
regime over which the 1/3 plateau (i.e.~the state of up-up-down spin order) in 
the magnetization is stabilized constitutes a quantity rather sensitive to the 
ratio $\Jzz'/\Jxx'$ \cite{Yamamoto2014,Yamamoto2019}. This plateau is evident 
in the data of Ref.~\cite{Xing2019}, although a lower temperature would assist 
a quantitative analysis. However, the relevant regime of parameter space is 
yet to be investigated theoretically for a field applied in the $ab$ plane of 
the system. Finally, while we cannot exclude terms of $J^{++}$ and $J^{+z}$ 
type from the spin Hamiltonian, the accuracy of our fit indicates that the 
effect of any missing terms is extremely small in \cybs; either they are 
genuinely small or they are relevant only at the lowest temperatures where
a treatment beyond the present mean-field level would be required.

On the materials side, it is also fair to say that the continuing lack of 
experimental signatures for QSL ground states can be blamed on two primary 
issues, namely the scarcity of candidate materials and the paucity of 
measurable physical quantities offering unambiguous signatures or predictors 
for QSL properties (such as fractional quantum numbers and nonlocal 
entanglement). To address the first of these, our analysis provides a 
definitive guide to the field-induced physics of highly spin-anisotropic 
systems such as the Yb-delafossites, and hence to the regions of parameter 
space where competing energy scales establish an environment conducive to 
the occurrence of exotic spin states. Although we cannot solve the second 
issue, we can provide a comprehensive understanding of the single-ion energy 
spectrum that identifies the extent to which a given material replicates a 
target pseudospin-1/2 model, thereby streamlining the experimental search 
for QSL fingerprints by available experimental methods. 

\section{Summary} 
\label{ss}

We have investigated the anisotropic magnetic response of an insulating 
$4f$ electronic system by measuring two key thermodynamic quantities, the 
magnetic susceptibility in the low-field limit and the magnetotropic 
coefficients over very wide field and temperature ranges (up to $\muh = 
60$ T and $T = 70 $ K). We have shown that the anisotropies in both 
quantities can be formulated within a set of anisotropic van Vleck (VV) 
coefficients, which arise as the second-order perturbative corrections 
of the Zeeman interaction to the zero-field crystal electric field (CEF) 
spectrum. This leads to the essential finding that the VV coefficients 
constitute independent physical quantities that describe the crucial 
magnetic properties of $4f$ spin systems across the full range of 
applied fields and extant anisotropies. A proper account of the 
ground-state VV coefficients is indispensable for an accurate and 
unambiguous determination of the microscopic parameters governing the 
CEF Hamiltonian, a process for which otherwise few routes are known to 
date. The VV coefficients fulfill the vital function of unifying the 
low-field, low-temperature magnetic susceptibility with the high-field 
magnetotropic coefficients and CEF levels, and in this sense their role 
as stand-alone physical quantities allowing a full interpretation of 
magnetic anisotropies has not been appreciated before.  

Our experimental results highlight the value of the resonant torsion 
magnetometry (RTM) method, which is accurate and profoundly powerful
in terms of the parameter ranges it accesses. The magnetotropic 
coefficients we extract over these broad field and temperature ranges 
play the key role in obtaining a unique set of Stevens operators 
describing the microscopic CEF Hamiltonian with unprecedent fidelity. 
We reiterate that a fitting analysis must provide complete consistency 
from zero to high field and at all relevant temperatures, and our fits 
meet this challenge. With the full CEF spectrum in hand, we can examine 
the validity of different and popular approximations that have been 
applied to many materials. Specifically, we identify the limits of a 
CW fit to the temperature-dependence of the magnetic susceptibility 
and the boundaries of the effective pseudospin-1/2 description for 
systems with a ground-state Kramers doublet. 

We have developed and applied our analysis for the material \cybs, 
which is a member of a family of Yb-delafossites displaying 
triangular-lattice geometry. Because the CEF levels of the Yb$^{3+}$ 
ion are four Kramers doublets, these compounds are leading candidates 
in the search for quantum spin-liquid (QSL) behavior, and indeed the 
full CEF spectrum we obtain up to high fields [Fig.~\ref{fig:spec}] 
reveals an intricate and anisotropic energy landscape amenable to 
unconventional magnetism. This spectrum allows one to construct a 
maximally informed pseudospin-1/2 model for the low-energy physics 
of the system, and within a minimal XXZ spin Hamiltonian we conclude 
that \cybs~is a strongly XY triangular antiferromagnet. While we 
await further experimental confirmation of this result, we note 
again that our analysis is applicable to a wide range of $4f$ 
materials with complex CEF spectra and especially with ground-state 
doublets allowing an effective spin-1/2 description, which should 
expand significantly the scope of the search for QSL phases.

\begin{acknowledgments}
We thank M. Mourigal, S. Nikitin, and K. Ross for helpful discussions. 
We are grateful to M. Chan and A. Shekhter for technical assistance 
with our pulsed-field measurements.
Experimental work at the University of Colorado Boulder was supported by 
Award No.~DE-SC0021377 of the U.S. Department of Energy (DOE), Basic Energy 
Sciences (BES), Materials Sciences and Engineering Division (MSE). 
Theoretical work at the University of Colorado Boulder was supported by  
Award No.~DE-SC0014415 of the U.S. DOE, BES, MSE. Work at Oak Ridge National 
Laboratory (ORNL) was supported by the U.S. DOE, BES, MSE. 
A portion of this work was performed at the National High Magnetic Field 
Laboratory, in the Pulsed Field Facility at Los Alamos National Laboratory 
(LANL), which is supported by National Science Foundation Cooperative 
Agreement No.~DMR-1644779, the State of Florida, and the U.S. DOE. 
The publication of this article was funded partially
by the University of Colorado Boulder Libraries Open Access Fund.
\end{acknowledgments}

\begin{appendix}

\section {Exact form of magnetic susceptibility } 
\label{app_exact}

From the $N$-particle partition function calculated with Eq.~(\ref{eq:HMF}), 
we derive analytical formulas for the low-field magnetic susceptibilities of 
our model using the thermodynamic relation 
\begin{widetext}
\begin{equation}
\chi_{ab(c)}(T) = - \frac{1}{\mu_0} \dfrac{\partial^2 F}{\partial H^2} \Big 
|_{H_{ab,c}\rightarrow 0} = \frac{N \mu_0 \mu_B^2 g_J^2}{k_B} \left[ \dfrac{T 
\displaystyle\sum_{n=0}^3 e^{-\beta E_n}}{\displaystyle\sum_{n=0}^3 \left( 
|\langle n_+ |\hat{J}_{x(z)}|n_{-(+)} \rangle|^2 - \alpha_{\perp(z)}^{n} 
\dfrac{2k_BT}{\mu_0^2 \mu_B^2 g_J^2} \right) e^{-\beta E_n} } + \Theta^{\rm 
CW}_{\perp(z)} \right]^{-1} \!\!\!\! , 
\label{eq:chifull}
\end{equation} 
\end{widetext}
where $\beta = 1/k_BT$, $\Theta^{\rm CW}_{\perp(z)} = q \mathcal{J}_{\perp(z)}/k_B$ 
as in Eq.~(\ref{eq:vvfit}), and $\alpha_{i}^n$ are the VV coefficients defined 
in Eq.~(\ref{eq:vvcoeff}). The form of Eq.~(\ref{eq:chifull}) is exact within 
the mean-field approximation, and by considering only $n = 0$ (the ground-state 
doublet) it reduces to Eq.~(\ref{eq:vvfit}).

\section{Comparison to the Curie-Weiss magnetic susceptibility of a 
spin-1/2 system}
\label{app_spinhalf}

To illustrate how the VV coefficients alter even the low-$T$ limit of the 
full susceptibility expression, where only the lowest Kramers doublet is 
relevant, we repeat Eq.~(\ref{eq:vvfit}) in the form
\begin{equation}
\! \chi_{ab(c)}^{-1} \! = \dfrac{k_B}{N \mu_0 \mu_B^2 g_J^2} \!\! \left[ 
\frac{T}{\! \left( \! \dfrac{g_{\perp(z)}}{2g_J} \!\! \right)^2 \!\!\! - \! 
\frac{2k_B \alpha_{\perp(z)}^0}{\mu_0^2 \mu_B^2 g_J^2} T} + \Theta^{\rm CW}_{\perp(z)} 
\! \right] \!.\!\!\!
\end{equation}
Only when the VV term is negligible does this expression reduce to the 
familiar CW form, 
\begin{equation}
\chi_{ab(c)}^{\rm CW} = \dfrac{N \mu_0 \mu_B^2 g_{\perp(z)}^2}{4k_B} \left( T
 + \widetilde \Theta_{\perp(z)}^{\rm CW} \right)^{-1} \!\!\!\! ,
\end{equation}
with a CW temperature for the effective $S = 1/2$ model given by 
$q \mathcal{J'}_{\perp(z)}/4 k_B$, where the parameters $\mathcal{J'}_{\perp}$ 
and $\mathcal{J'}_{z}$ are defined in App.~\ref{app_proj} below.

\section{Effective spin-1/2 projection}
\label{app_proj}

At the lowest temperatures, where the physics is dominated by the properties 
of the lowest Kramers doublet, $\{|0_\pm\rangle\}$, it is helpful to formulate 
a pseudospin-1/2 model in terms of effective spin-1/2 operators with the action
\begin{align}
\hat{S}_x & = {\textstyle \frac12} \left( |0_+ \rangle \langle 0_- | + |0_- 
\rangle \langle 0_+| \right) , \\
\hat{S}_z & = {\textstyle \frac12} \left( |0_+ \rangle \langle 0_+ | - |0_- 
\rangle \langle 0_-| \right) .
\end{align}
The equivalent pseudospin-1/2 model for a Hamiltonian $\hat{\mathcal{H}}$ is 
obtained formally by projecting onto the lowest Kramers-doublet subspace 
using the projection operator
\begin{equation}
\hat{P}_0 \equiv |0_+\rangle\langle 0_+| + |0_-\rangle\langle 0_-|.
\end{equation}
For the full $J = 7/2$ Hamiltonian in the form
\begin{equation}
\hat{\mathcal{H}} = \sum_{\langle i,j\rangle} \left( \mathcal{J}_{\perp}\hat{J}_{i,x}
\hat{J}_{j,x} + \mathcal{J}_{z}\hat{J}_{i,z}\hat{J}_{j,z} \right) + \mu_0 \mu_B g_J
\mathbf{H} \cdot \sum_{i} \hat{\mathbf{J}}_i,
\end{equation}
the equivalent pseudospin-1/2 model is 
\begin{align*}
\hat{\mathcal{H}}_{\rm PS} &= \hat{P}_0 \hat{\mathcal{H}} \hat{P}_0,
\numberthis \\ & =  \sum_{\langle i,j\rangle} \left( \mathcal{J}_{\perp}' 
\hat{S}_{i,x} \hat{S}_{j,x} + \mathcal{J}_{z}' \hat{S}_{i,z} \hat{S}_{j,z} \right) 
\\ & \qquad + \mu_0 \mu_B \sum_{i} \left( g_{\perp} H_\perp \hat{S}_{i,x} + g_{z} 
H_z \hat{S}_{i,z}\right), \numberthis
\end{align*}
from which it follows that the equivalent $g$-factors and exchange constants 
of the pseudospin-1/2 model are related to $J = 7/2$ operator matrix elements 
and exchange constants by the expressions
\begin{align}
g_{\perp(z)} & = 2g_J|\langle 0_{\pm}|\hat{J}_{x(z)}|0_{\mp(\pm)}\rangle|,\\
\mathcal{J}_{\perp(z)}' & = \left(\dfrac{g_{\perp(z)}}{g_J} \right)^2
\mathcal{J}_{\perp(z)}. 
\end{align}
Applying these relations using $g_J = 8/7$ gives the $g$-factors $g_\perp = 
3.77$ and $g_z = 1.76$ shown in the second line of Table 2, and using the 
results $\Jxx = 0.54 \pm 0.01$ K and $\Jzz = 0.61 \pm 0.01$ K obtained in 
Sec.~\ref{sechi}B for the $J = 7/2$ system gives the values $\Jxx' = 5.12$ 
K ($\simeq 0.44$ meV) and $\Jzz' = 0.84$ K ($\simeq 0.07$ meV) for the 
interaction parameters of the pseudospin-1/2 model.

\section{Magnetotropic Coefficients} 
\label{app_rtm}

In an anisotropic magnetic material, a magnetic-field-angle-dependent 
contribution to the Helmholtz free energy,
\begin{align}
F(T,H,\vartheta) = -\frac{1}{\beta}{\rm log}\Tr\left[e^{-\beta\hat{\mathcal{H}}
(H,\vartheta)}\right], \label{eqn:Fang}
\end{align}
can be modelled using the explicit field-dependence of the microscopic spin 
Hamiltonian, $\hat{\mathcal{H}}(H,\vartheta)$. In Sec.~\ref{smm} we model 
this at the mean-field level, using $\hat{\mathcal{H}}_{\rm MF}^{\rm sg}$ 
[Eq.~(\ref{eq:HMF})]. RTM is a direct probe of the curvature of the 
magnetic free energy with respect to changes in the angle at which the 
field is applied, and this curvature is defined as the magnetotropic 
coefficient, 
\begin{equation}
k (H,\vartheta) = \frac{\partial^2 F}{\partial \vartheta^2} =
 -\frac{\partial\tau(H,\vartheta)}{\partial \vartheta}, 
\label{eq:magneotropic}
\end{equation} 
where $\tau (H, \vartheta) = \mathbf M\times \mathbf H$ is the magnetic 
torque and $\vartheta$ is the angular direction of $H$ measured in the 
plane of vibration of the sample. For the system of a piezoresistive 
cantilever with a tip-mounted sample, the angle-dependent magnetotropic 
coefficient at finite field produces the proportional frequency shift 
specified in Eq.~(\ref{eq:fs}) \cite{Modic2018,Modic2020}.
For the cases considered here, we define the total vector magnetization in 
a finite field with components $H_{\perp} = H \sin \theta$ and $H_z = H \cos 
\theta$ in terms of its component magnetizations as 
\begin{equation}
\mathbf{M}(\mathbf{H}) = M_{ab}(H_{\perp}, H_z)\hat{\mathbf{x}} + M_{c}
(H_\perp, H_z)\hat{\mathbf{z}}.
\end{equation}
The torque as function of angle then takes the explicit form
\begin{equation}
\tau(\theta) = \mathbf{M} \times \mathbf{H} = M_{ab} H \cos \theta
 - M_{c} H \sin \theta.
\end{equation}
In addition to the conventional susceptibilities,
\begin{eqnarray}
\chi_{ab} & = & \lim_{h\to0} \dfrac{M_{ab}(H_{\perp} = h, H_z = 0)}{h} 
\nonumber \\ \chi_{c} & = & \lim_{h \to 0} \dfrac{M_{c} (0,h)}{h}, 
\nonumber
\end{eqnarray}
one may define the transverse susceptibilities at finite field, for a 
magnetization induced by a small field $h$ applied perpendicular to an 
arbitrarily large (but finite) field $H$, as    
\begin{align}
\chi_{ab}^T(H) & = \lim_{h\to 0} \dfrac{M_{ab} (H_{\perp} \! = \! h, 
H_z \! = \! H) - M_{ab}(0,H)} {h}, \\ \chi_c^T(H) & = \lim_{h \to 0} 
\dfrac{M_c(H,h) - M_c(H,0)}{h}.  
\end{align}
With these definitions we express the magnetotropic coefficients at the 
high-symmetry angles as $k_{ab}(H) = k(H, \theta = \pi/2)$ and $k_{c}(H) = 
k(H, \theta = 0)$, with 
\begin{align}
k_{ab}(H)  & = H M_{ab}(H,0) - H^2 \chi_{c}^T (H),\\
k_{c}(H) & = H M_{c} (0,H) - H^2 \chi_{ab}^T (H).
\end{align}
In the low-field limit, the transverse susceptibilities are in fact 
identical to their zero-field values, so that $k_{ab} = -k_c = (\chi_{ab}
 - \chi_c) H^2$. The low-field expressions are no longer valid when the 
magnetization ceases to be $H$-linear, necessitating a more detailed 
description in terms of Eqs.~(\ref{eqn:Fang}) and (\ref{eq:magneotropic}). 
From the expressions at low and high fields, one may interpret the 
magnetotropic coefficient at a given applied-field angle, $\theta$, as 
the energy difference between the real magnetostatic potential energy 
at that angle and the ``naive'' energy cost of magnetizing the sample 
in a completely perpendicular direction based on a linear extrapolation 
of the transverse susceptibility.  

\end{appendix}


\begin{thebibliography}{80}
\bibitem [{DOn()}]{DOnote}%
  \BibitemOpen
  \href@noop {} {}\bibinfo {note} {This can be understood by examining the
  eigenvalues of the threefold rotation operator, $\exp(i\frac{2\pi
  J_z}{3})$.}\BibitemShut {Stop}%
\end{thebibliography}
%

\end{document}